\documentclass[aps,prd,eqsecnum,notitlepage,nofootinbib]{revtex4}
\usepackage[centertags]{amsmath}
\usepackage{amssymb}
\usepackage{amsfonts,mathrsfs}
\usepackage{hyperref}	
\usepackage{graphicx}
\usepackage{dcolumn}
\usepackage{bm}
\usepackage{tabularx}
\usepackage{array}
\usepackage{float}
\usepackage{braket}
\usepackage{color}
\usepackage{enumerate}

\newcommand{\be}{\begin{equation}}
\newcommand{\ee}{\end{equation}}
\newcommand{\bea}{\begin{eqnarray}}
\newcommand{\eea}{\end{eqnarray}}
\newcommand{\bes}{\begin{subequations}}
\newcommand{\ese}{\end{subequations}}
\usepackage{tikz}
\usetikzlibrary{positioning,shapes}
\usepackage{subfigure} 
\usepackage{enumitem}

\newlist{inlineroman}{enumerate*}{1}
\setlist[inlineroman]{itemjoin*={{, and }},afterlabel=~,label=\roman*.}

\newlist{Inlineroman}{enumerate*}{1}
\setlist[Inlineroman]{itemjoin*={{, and }},afterlabel=~,label=\Roman*.}

\newcommand{\InlineRom}[1]{
\begin{Inlineroman}
#1
\end{Inlineroman}
}

\begin{document}

\title{Estimation of two-qubit  interactions through channels with environment assistance}

\author{MILAJIGULI REXITI
}

\address{School of Advanced Studies, University of Camerino, 62032 Camerino, Italy\\
milajiguli.milajiguli@unicam.it}

\author{STEFANO MANCINI}

\address{School of Sience and Technology, University of Camerino, 62032 Camerino, Italy\\
INFN-Sezione di Perugia, I-06123 Perugia, Italy\\
stefano.mancini@unicam.it}

\begin{abstract}
We consider the estimation of two-qubit interactions when initial states of both qubit can be locally controlled, while the final state of only one qubit can be measured. This amounts to realize a model of quantum channel communication with environment assistance. In such a framework the unitaries' � parameters space results a tetrahedron in $\mathbb{R}^3$. On its edges the problem, becoming of single parameter estimation, can be exactly solved and we derive optimal probe states and estimators. Our results show that the possibility of environment assistance is always beneficial, while the usage of entanglement at channels' input is not.
\end{abstract}

\maketitle

\markboth{M.Milajiguli, S.Mancini}
{Estimation of two-qubit  interactions through channels with environment assistance}

\section{Introduction}	

Recently quantum estimation theory recieved renewed attention after seminal work of Helstrom\cite{Helstrom}, mainly due to the advent of quantum technologies \cite{Paris}.

Quantum estimation aims at devising optimal strategies to determine the value of quantities that are not directly accessible/observable. These quantities are parameters inherent to a physical transformation/process. Then the strategy resorts to the preparation of probe states that undergone the transformation and the realization of a Probability Operator Valued Measure (POVM) on the resulting transformed states. Hence, a double optimization procedure is often involved.

Unitaries are transformation employed in ideal description of physical processes. This double optimization problem has been studied in the context of estimation of ${\rm SU}(d)$ unitary 
operations \cite{dep1}. Moreover, in such a context, the usage of entanglement at input has been shown to improve the estimation accuracy\cite{dep}.

In realistic settings quantum channels should be evoked instead of unitaries for describing physical processes. A quantum channel is a Completely Positive and Trace Preserving (CPTP) map on the set of states over a Hilbert space \cite{mark}. The issue of estimating a quantum channel has been discussed in the literature\cite {fujiwara1,dep3,dep4} . Also there the usefulness of entanglement has been pointed out. Thanks to Stinespring dilation \cite{Stinespring}, a quantum channel can be viewed as coming from a unitary between system and environment after tracing out the latter. Hence estimation of such unitary can be attempted to some extent by looking at the quantum channel's transformation.

We shall address here the issue of estimating a parametrized family of system-environment interaction unitaries by analysing the channels arising from it. To this end we shall employ a model of quantum channel with environment assistance recently introduced in Ref.\cite{sid}. More specifically we consider two qubit unitaries when initial states of both qubit can be locally controlled, while the final state of only one qubit can be measured. In such a framework only entangling unitaries become relevant\footnote{With local unitaries the problem reduces to the estimation of a single qubit unitaries.} and they form a subset of ${\rm SU} (4) $ that can be characterized by three real parameters \cite{nonlocal gates}. Actually the parameters space results a tetrahedron in $\mathbb{R}^3$.  On its edges the problem, becoming of single parameter estimation, will be exactly solved by minimizing a suitable cost function averaged over all possible values of the parameter. We shall hence derive optimal probe states and estimators (POVM). Our results show that the possibility of environment assistance is always beneficial, while the usage of entanglement at channels input is not.

\section{Environment Assistance Model}

\begin{figure}[ht]
\begin{center}
\begin{tikzpicture}[scale=0.5]
\draw[thick] (2,4) -- (10,4);
\draw[thick] (0,0) -- (4,0); 
\draw[thick] (2,2) -- (4,2); 
\draw[thick] (0,3) -- (2,4);
\draw[thick] (0,3) -- (2,2);
\draw[thick] (8,2) -- (10,2);
\draw[thick] (8,0) -- (10,0);
\draw[thick] (4,-1) rectangle (8,3);
\node[cloud, cloud puffs=15.7, cloud ignores aspect, minimum width=0.3cm, minimum height=1cm, align=center, draw] (cloud) at (10.5,0){}; 
\node[left] at (0,0){$\eta$};
\node[below] at (2,0){$E$};
\node[below] at (9,2){$B$};
\node[left] at (0,3){$\phi_{in}$}; 
\node[right] at (10,3){$\phi_{out}$}; 
\node[below] at (2,2){$A$}; 
\node[above] at (6,4){$R$}; 
\node[below] at (9,0){$F$};
\draw (6,1) node[font = \fontsize{40}{42}\sffamily\bfseries]{$U$};
\end{tikzpicture}
\end{center}
 \caption{ Quantum channel with environment assistance model. The quantum  channel between input system $A$ and output system $B$ arises from a unitary $U$ by tracing out $F$. The input system $A$ can be entangled with a reference system $R$. The environment system $E$ can be controlled by the helper.} \label{model}
\end{figure}
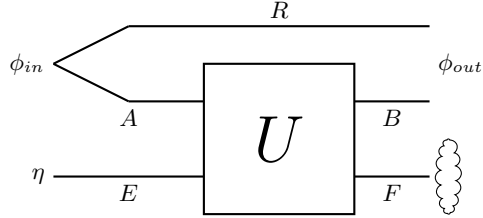

Every quantum channel can be viewed as arising from the unitary interaction of a system with its environment \cite{Stinespring}. The resulting entanglement between system and environment is lost when the environment is `traced out' thereby, destroying the purity of the signal states and introducing noise into the system.
A communication model with environment assistance conceives a third party, other than sender and receiver, who can control the environment input system as sketched in Fig.\ref{model}  \cite{sid}. 
In such a context we will study the estimation of two-qubit unitaries representing the dilation of qubit channels into a single qubit environment.  

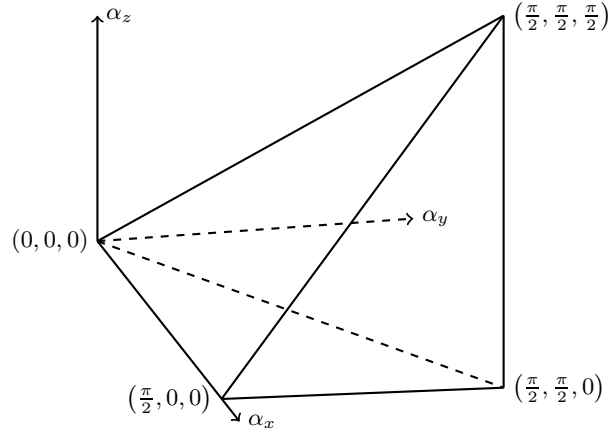
\begin{figure}[ht]
\begin{center}
\begin{tikzpicture}[scale=0.3]
\draw[thick,dashed][->] (0,0) -- (14,1); 
\node[right] at (14,1){$\alpha_y$};
\draw[thick][->] (0,0) -- (0,10); 
\node[right] at (0,10){$\alpha_z$};
\draw[thick][->] (0,0) -- (6.3,-8); 
\node[right] at (6.3,-8){$\alpha_x$};
\draw[thick,dashed] (0,0) -- (18,-6.5); 
\draw[thick] (5.5,-7) -- (18,-6.5); 
\draw[thick] (18,10) -- (18,-6.5); 
\draw[thick] (18,10) -- (5.5,-7); 
\draw[thick] (18,10) -- (0,0); 
\node[left] at(0,0) {$\left(0,0,0 \right)$};
\node[right] at(18,10) {$\left(\frac{\pi}{2},\frac{\pi}{2},\frac{\pi}{2} \right)$};
\node[right] at(18,-6.5) {$\left(\frac{\pi}{2},\frac{\pi}{2},0 \right)$};
\node[left] at(5.5,-7) {$\left(\frac{\pi}{2},0,0 \right)$};
\end{tikzpicture}
\end{center}
\caption{Tetrahedron representing the parameters space of two-qubit unitaries.}
\label{tetrahedron}
\end{figure}

 Specifically, letting $A$ denote the system qubit and $E$ the environment qubit,
given a unitary $U_{AE}$ that entangles system and environment,
 the action of the channel $\mathscr{N}$ on $A$ is obtained as 
\begin{equation}
\mathscr{N}\left(\phi_{in}\right)={\rm Tr}_F \left[\left(I_R\otimes U_{AE}\right)\left(\phi_{in} \otimes \eta \right)\left( U_{AE}^\dag \otimes  I_R \right)\right],
\end{equation}
where  $\phi_{in}=\ket \phi_{in}\bra \phi $ is the input state for the system $A$ possibly entangled with 
reference system $R$, and $\eta= \ket \eta \bra\eta$ is the state of the environment $E$. 
The unitary $U_{AE}$ acting on two qubits Hilbert space $\mathscr{H}_{AE}\simeq \mathbb{C}^2\otimes\mathbb{C}^2$ is a member of ${\rm SU}(2\times 2)$. However among all possible members of such a group, those entangling system and environment,
can be written as \cite{nonlocal gates}:
\begin{equation}
U_{AE}=\sum_{k=1}^{4} e^{-i \lambda_k} \ket{\Lambda_k}\bra{\Lambda_k}=\sum_{k=1}^{4} (\cos \lambda_k -i \sin \lambda_k )\ket{\Lambda_k}\bra{\Lambda_k}, 
\label{un}
\end{equation}
where $ \ket{\Lambda_k} $ are the so called magic basis states:
\begin{align}
\ket{\Lambda_1}=\frac {1}{\sqrt 2}(\ket{0_A}\ket{0_E}+\ket{1_A}\ket{1_E}), \hspace{10mm} 
\ket{\Lambda_2}=\frac {-i}{\sqrt 2}(\ket{0_A}\ket{0_E}-\ket{1_A}\ket{1_E}),
\notag\\ 
\ket{\Lambda_3}=\frac {1}{\sqrt 2}(\ket{0_A}\ket{1_E}-\ket{1_A}\ket{0_E}),\hspace{10mm} 
\ket{\Lambda_4}=\frac {-i}{\sqrt 2}(\ket{0_A}\ket{1_E}+\ket{1_A}\ket{0_E}),
\end{align}
and the eigenvalues $\lambda_k$  are
\begin{align}
\lambda_1=&\frac{\alpha_x-\alpha_y+\alpha_z}{2},   \hspace{12mm} \lambda_2=\frac{-\alpha_x+\alpha_y+\alpha_z}{2},  \notag\\
\lambda_3=&\frac{-\alpha_x-\alpha_y-\alpha_z}{2},  \hspace{9mm}
\lambda_4=\frac{\alpha_x+\alpha_y-\alpha_z}{2},
\end{align}
with 
\begin{equation}
\frac{\pi}{2}\geq \alpha_x \geq \alpha_y \geq \alpha_z \geq 0.
\end{equation}
In such a way the set of unitaries we are going to consider can be parametrized by 3 real parameters instead of  the usual 15. Hence the parameter space
\begin{equation}\label{spaceS}
\mathscr{S}=\left\{(\alpha_x, \alpha_y, \alpha_z): \frac{\pi}{2}\geq \alpha_x \geq \alpha_y \geq \alpha_z \geq 0\right\},
\end{equation}
describes all two-qubit unitaries up to local basis choice and complex conjugation. The parameters $\{\alpha_x,\alpha_y,\alpha_z\}$ are $\frac{\pi}{2}$-periodic and symmetric around $\frac{\pi}{4}$.  They form a tetrahedron with vertices 
$(0,0,0), \left(\frac{\pi}{2},0,0\right), \left(\frac{\pi}{2},\frac{\pi}{2},0\right)$ and$\left(\frac{\pi}{2},\frac{\pi}{2},\frac{\pi}{2}\right)$ shown in Fig.\ref{tetrahedron}. Familiar two-qubits unitaries (gates) can easily be identified within this parameter space: for instance, $(0,0,0)$ represents the identity $I$, $\left(\frac{\pi}{2},0,0\right)$ represents the CNOT, $ \left(\frac{\pi}{2},\frac{\pi}{2},0\right)$ the DCNOT (double controlled not), and $\left(\frac{\pi}{2},\frac{\pi}{2},\frac{\pi}{2}\right)$ the SWAP gate, respectively. 

 The matrix representation of the unitaries \eqref{un} in the canonical basis $\{\ket{0_A}\ket{0_E},\ket{0_A}\ket{1_E}, \ket{1_A}\ket{0_E}, \ket{1_A}\ket{1_E}\}$ reads:
\begin{equation}
U_{AE}=\left(
\begin{array}{cccc}
 e^{-\frac{1}{2} i \alpha_z} \cos \left(\frac{\alpha_x-\alpha_y}{2}\right) & 0 & 0 & -i e^{-\frac{1}{2} i \alpha_z} \sin \left(\frac{\alpha_x-\alpha_y}{2}\right) \\
 0 & e^{\frac{i }{2}\alpha_z} \cos \left(\frac{\alpha_x+\alpha_y}{2}\right) & -i e^{\frac{i }{2}\alpha_z} \sin \left(\frac{\alpha_x+\alpha_y}{2}\right) & 0 \\
 0 & -i e^{\frac{i }{2}\alpha_z} \sin \left(\frac{\alpha_x+\alpha_y}{2}\right) & e^{\frac{i }{2}\alpha_z} \cos \left(\frac{\alpha_x+\alpha_y}{2}\right) & 0 \\
 -i e^{-\frac{1}{2} i \alpha_z} \sin \left(\frac{\alpha_x-\alpha_y}{2}\right) & 0 & 0 & e^{-\frac{1}{2} i \alpha_z} \cos \left(\frac{\alpha_x-\alpha_y}{2}\right)
\end{array}
\right). \label{U}
\end{equation}

\section{Parameter Estimation}

Although the space of parameters \eqref{spaceS} is the entire tetrahedron, we shall focus 
 here on the estimation of parameters lying along the edges of tetrahedron,
 because in such a case the problem turns out to be analytically solvable.
 
By referring to Fig.\ref{model} we aim at finding the optimal POVM to apply to the output $B$ and the optimal input state for the system $RA$ as well as for the environment $E$, such that we can estimate the unknown parameter on the edge of tetrahedron with the smallest cost.
This amounts to solving a double optimization problem.

\subsection{Input (Probe) and Output States}

Let the input state for the global system (probe and environment) be
\begin{equation}
|\Phi\rangle_{in}=|\phi_{in}\rangle|\eta\rangle=
\left(\sqrt{x}\ket {0_R} \ket {0_A}+e^{i \varphi_1}\sqrt{1-x}\ket{ 1_R} \ket {1_A}\right) \left(\sqrt{t}\ket {0_E} +e^{i \varphi_2}\sqrt{1-t}\ket {1_E} \right), 
\label{input}
\end{equation}
where $0\le x,t\le 1$, $\varphi_{1,2}\in[0,2\pi]$. 
Assuming that entanglement with reference system can be exploited, 
hence to be able to also measure the system $R$ (besides the system $B$),
the output state to which apply a POVM will be
\begin{eqnarray}
\rho={\rm Tr}_F |\Phi\rangle_{out}\langle \Phi |,
\end{eqnarray}
where 
\begin{eqnarray}
\ket \Phi_{out}= \left(I_R\otimes U_{AE}\right) \ket\Phi_{in},
\end{eqnarray}
with $U$ given by \eqref{U}. Taking into account \eqref{U} and \eqref{input},
the matrix form of the output state in the canonical basis reads:
\begin{equation}
 \left(
\begin{array}{cccc}
 \frac{ x}{2} (1+(2 t-1) \zeta+\xi) & \sigma_1 & \sigma_2 & \sigma_3 \\
 \sigma_1^* & \frac{x}{2}  (1-(2 t-1) \zeta-\xi) & \sigma_4 & -\sigma_2 \\
  \sigma_2^* &  \sigma_4^* & \frac{x-1}{2} (\xi+(1-2 t) \zeta-1) & \sigma_5 \\
  \sigma_3^* &  -\sigma_2^* &  \sigma_5^* & \frac{x-1}{2} (-\xi-(1-2 t) \zeta-1) \\
\end{array}
\right), 
\label{output}
\end{equation}
where ${}^*$ denotes complex conjugation and
\begin{eqnarray*}
\xi &=&\cos \alpha_x \cos \alpha_y, \\
\zeta &=&\sin \alpha_x \sin \alpha_y,\\
 \sigma_1 &=& \sqrt{(1-t) t} \, x\,  (\sin \alpha_y \sin (\alpha_z+\phi_2)+i \sin \alpha_x \cos (\alpha_z+\phi_2)),\\
 \sigma_2 &=& \frac{1}{2} i \sqrt{(1-t) t} \sqrt{(1-x) x} \, e^{-i (\phi_1+\phi_2)} \left(\sin (\alpha_x-\alpha_y)+e^{2 i \phi_2} \sin (\alpha_x+\alpha_y)\right),\\
 \sigma_3 &=& \frac{1}{2} e^{-i \phi_1} \sqrt{(1-x) x} (\cos \alpha_x+\cos \alpha_y) (\cos \alpha_z+i (1-2 t) \sin \alpha_z), \\
 \sigma_4 &=& \frac{1}{2} e^{-i \phi_1} \sqrt{(1-x) x} (\cos \alpha_y-\cos \alpha_x) (\cos \alpha_z+i (1-2 t) \sin \alpha_z), \\
 \sigma_5 &=&- i \sqrt{(1-t) t} (1-x) (\sin \alpha_x \cos (\alpha_z-\phi_2)+i \sin \alpha_y \sin (\alpha_z-\phi_2)).
\end{eqnarray*}

\subsection{Single Parameter Estimation with Quadratic Cost Function}\label{sec:quadratic}
Suppose we have an unknown parameter $\alpha$ in $\mathscr{I}\subset\mathscr{S}$ to be estimated 
through a quantum measurement on the output state $\rho(\alpha)$. 
We consider a POVM  $\Pi(\hat\alpha)$, where $\Pi(\hat\alpha)\ge 0$ and  
$\int_\mathscr{I}{\Pi}(\hat{\alpha})d \hat{\alpha}={I}$, whose outcome $\hat{\alpha}$ will be an estimate of 
$\alpha$. 
The conditional probability to have the estimated value $\hat\alpha$ given the parameter value $\alpha$ is
 \begin{equation}
p\left( \hat \alpha| \alpha \right)={\rm Tr}[{\Pi}(\hat{\alpha}){\rho}(\alpha)].
\end{equation}
By means of that it is possible to define the quadratic cost function for the taken estimation strategy  
\begin{equation}
C:=\int_\mathscr{I}\left(\alpha-\hat{\alpha}\right)^2 {\rm Tr}[{\Pi}(\hat{\alpha}){\rho}(\alpha)]d\hat{\alpha}.
\end{equation}
Since the value taken by the parameter is not known, we have to assume no a priori knowledge about 
$\alpha$, that is an a priori flat probability distribution  $p(\alpha)$. 
Then we want to minimize the average of the quadratic cost function:
\begin{equation}
\bar{C}=\int_\mathscr{I}\int_\mathscr{I}(\alpha-\hat{\alpha})^2  {\rm Tr}[{\Pi}(\hat{\alpha}){\rho}(\alpha)]d\hat{\alpha} p(\alpha)d\alpha
\label{cost}
\end{equation} 
It is customary to introduce the risk operator
\begin{equation}
{W}(\hat\alpha):=\int_\mathscr{I} p(\alpha) (\alpha-\hat\alpha)^2{\rho}(\alpha)d\alpha
={W}^{(2)}-2\hat\alpha {W}^{(1)}+\hat\alpha^2 {W}^{(0)}, 
\label {risk}
\end{equation} 
where 
\begin{equation}
{W}^{(k)}:= \int_\mathscr{I} p(\alpha)\alpha^k {\rho}(\alpha)d\alpha, \quad k=0,1,2. 
\label{riske}
\end{equation}
{ Then the average cost function \eqref{cost} becomes
\begin{equation}
\bar{C}={\rm Tr}  \int_\mathscr{I} \Pi(\hat\alpha){W}(\hat\alpha) d\hat{\alpha}.
\label{cost1}
\end{equation}
It is known that the POVM $\Pi(\hat\alpha)$ minimizing $\bar{C}$ can be found through 
the solution $\Theta $ of the following operator equation \cite{Helstrom} 
\begin{equation}
\Theta  W^{(0)}+ W^{(0)} \Theta=2 W^{(1)}.
\label{eqM}
\end{equation} 
When the solution exists as an Hermitian operator
we can write its spectral decomposition 
\begin{equation}
 \Theta=\sum_{i=1}^4\theta_i\ket{ \theta_i} \bra {\theta_i}.
\end{equation}
Then, the optimal POVM is given by
\begin{equation}
\Pi(\hat\alpha)=\sum_{i=1}^4 \delta(\hat\alpha-\theta_i)\ket{ \theta_i} \bra {\theta_i}. 
\label{povm}
\end{equation}
This implies that the measurement has four outputs at most and we estimate the parameter as one of four 
$\theta_i$'s. 
The minimum average cost results
\begin{equation}
\min \bar C={\rm Tr}\left( W^{(2)}-\Theta W^{(1)} \right).
\label{minbarC}
\end{equation}
}

\section{Results}
{
By referring to Fig.\ref{tetrahedron} the edges of tetrahedron are given by:
\InlineRom{\item $ \alpha_x=\frac{\pi}{2}, \alpha_y=\frac{\pi}{2} $ ; \item $ \alpha_y=0, \alpha_z=0 $;
\item $\alpha_x=\alpha_y, \alpha_z=0  $ ; 
\item  $\alpha_x=\alpha_y=\alpha_z $  ;
\item $\alpha_x=\frac{\pi}{2}, \alpha_z=0 $
\item  $\alpha_x=\frac{\pi}{2},\alpha_y=\alpha_z $.}
Hence, each edge is characterized by a single parameter.
For each of them we calculate the minimum average cost \eqref{cost1} with generic input \eqref{input}, 
then we optimize over the input state, i.e., find the best probe and environment states. 
}


\subsection{Edge $ \alpha_x=\frac{\pi}{2}, \alpha_y=\frac{\pi}{2} $ }
\label{edge1}

{ We consider the output state \eqref{output} with $ \alpha_x=\frac{\pi}{2}, \alpha_y=\frac{\pi}{2} $, so that $\alpha_z\equiv\alpha$ becomes the parameter to estimate. 
Then, to solve the matrix equation \eqref{eqM} we vectorize it to be
\begin{equation}
(I_4 \otimes {W}^{(0)}+ {W}^{(0)^T} \otimes I_4)\left(Vec\, \Theta\right)=2\left(Vec\, {W}^{(1)}\right),
\end{equation}
where $I_4$ is $4\times 4 $ identity matrix and $Vec(\bullet)$ denotes vectorization of the matrix $\bullet$. In this form, the equation can be seen as a linear system of 16 equations. When the coefficient matrix $(I_4 \otimes {W}^{(0)}+ {W}^{(0)^T} \otimes I_4)$ is not singular,  \eqref{eqM} has unique solution. Hence, for $0<x,t<1$ 
we get}
\begin{equation}
\Theta=\left(
\begin{array}{cccc}
 \frac{\pi }{4} & \frac{(1-i)  e^{-i\varphi_2} (4-\pi ) \sqrt{(1-t) t} }{\pi } & 0 & 0 \\
 \frac{(1+i) e^{i\varphi_2} (4-\pi ) \sqrt{(1-t) t}}{\pi } & \frac{\pi }{4} & 0 & 0 \\
 0 & 0 & \frac{\pi }{4} & \frac{(1+i) e^{-i \varphi_2} (4 -\pi) \sqrt{(1-t) t}}{\pi } \\
 0 & 0 & \frac{(1-i) e^{i \varphi_2} (4-\pi ) \sqrt{(1-t) t}}{\pi } & \frac{\pi }{4} \\
\end{array}
\right).
\end{equation}
{Then, Eq.\eqref{minbarC} yields}
\begin{equation}
\bar{C}(x,t,\varphi_1, \varphi_2)=\frac{96 \pi ^2 (t-1) t-768 \pi  (t-1) t+1536 (t-1) t+\pi ^4}{48 \pi ^2},
\label{6}
\end{equation}
{which turns out to be only dependent on the environment state through $t$.
It attains the minimum }
\begin{equation}
\min\bar C= -\frac{1}{2}-\frac{8}{\pi ^2}+\frac{4}{\pi }+\frac{\pi ^2}{48} \approx 0.17,
\end{equation}
when $t=\frac{1}{2}$, i.e. with a superposition of canonical basis states for $E$.

{ Although $\bar{C}$ does not depend on $\varphi_1, \varphi_2,x$ the minimizing operator $\Theta$ does depend on $\varphi_2$ and hence the POVM. To derive it we can therefore set w.l.o.g. $\varphi_2=0$ and obtain}
\begin{eqnarray}
\ket \theta_1 =\left(
\begin{array}{c}
0 \\ 0 \\ \frac{1+i}{2}\\\frac{1}{\sqrt{2}}
\end{array}
\right), \quad
\ket \theta_2 =\left(
\begin{array}{c}
 \frac{1-i}{2} \\ \frac{1}{\sqrt{2}} \\0 \\ 0
\end{array}
\right),  \quad
\ket \theta_3 =\left(
\begin{array}{c}
0 \\ 0 \\ \frac{-1-i}{2}\\\frac{1}{\sqrt{2}}
\end{array}
\right), \quad
\ket \theta_4 =\left(
\begin{array}{c}
\frac{-1+i}{2} \\ \frac{1}{\sqrt{2}} \\ 0 \\ 
\end{array}
\right),
\end{eqnarray}
{ together with}
\begin{eqnarray}
\theta_1&=&\theta_2=\frac{2 \sqrt{2}}{\pi }+\frac{\pi }{4}-\frac{1}{\sqrt{2}},\\
\theta_3&=&\theta_4=-\frac{2 \sqrt{2}}{\pi }+\frac{\pi }{4}+\frac{1}{\sqrt{2}}.
\end{eqnarray}
When $x=0$,  the matrix  $(I_4 \otimes {W}^{(0)}+ {W}^{(0)^T} \otimes I_4)$ becomes singular and \eqref{eqM} has infinitely many solutions which can be summarized as
\begin{equation}
\Theta= \left(
\begin{array}{cccc}
 k_1 & k_2 & 0 & 0 \\
 k_2 & k_3 & 0 & 0 \\
 0 & 0 & \frac{\pi }{4} &\frac{(1+i) (4-\pi ) \sqrt{(1-t) t}}{\pi }  \\
 0 & 0 &\frac{(1-i) (4-\pi ) \sqrt{(1-t) t}}{\pi }& \frac{\pi }{4} \\
\end{array}
\right),
\end{equation}
where $k_1, k_2, k_3$ are arbitrary real constants satisfying
\begin{eqnarray}
& &0 \leq -\sqrt{(k_1^2- k_3)^2+4 k_2^2}+k_1+k_3 \leq \pi,  
\label{k1}\\
& & 0\leq \sqrt{(k_1^2- k_3)^2+4 k_2^2}+k_1+k_3 \leq \pi.
\label{k2}
\end{eqnarray}
These conditions are imposed by the fact that the parameter we are estimating must lie in the interval 
$\left[0, \frac{\pi}{2}\right]$ and hence must do the eigenvalues of minimizing operator.

{ The minimum average cost function results the same of Eq.\eqref{6}.
With $t=\frac{1}{2}$, and w.l.o.g. $\varphi_1=\varphi_2=0$, we obtain} 
\begin{eqnarray}
\ket \theta_1& =&\left(
\begin{array}{c}
\frac{sign(k_2)}{\sqrt{2}} \sqrt{1-\frac{k_1-k_3}{\sqrt{(k_1-k_3)^2+4 k_2^2}}}
 \\ \frac{\sqrt{2k_2^2}}{\sqrt{4k_2^2-(k_1-k_3)\left(\sqrt{(k_1-k_3)^2+4k_2^2}-k_1+k_3\right)}}\\0\\0
\end{array}
\right), \quad \notag \\
\ket \theta_2 &=&\left(
\begin{array}{c}
-\frac{sign(k_2)}{\sqrt{2}} \sqrt{1+\frac{k_1-k_3}{\sqrt{(k_1-k_3)^2+4 k_2^2}}}
 \\ \frac{\sqrt{2k_2^2}}{\sqrt{4k_2^2+(k_1-k_3)\left(\sqrt{(k_1-k_3)^2+4k_2^2}+k_1-k_3\right)}}\\0\\0
\end{array}
\right), \notag \\
\ket \theta_3 &=&\left(
\begin{array}{c}
0 \\ 0 \\ \frac{1}{2}+\frac{i}{2}\\\frac{1}{\sqrt{2}}
\end{array}
\right), \quad
\ket \theta_4 =\left(
\begin{array}{c}
0 \\ 0 \\ -\frac{1}{2}-\frac{i}{2} \\ \frac{1}{\sqrt{2}}
\end{array}
\right),
\end{eqnarray}
together with
\begin{eqnarray}
\theta_1&=&\frac{1}{2} \left(k_1 + k_3 - \sqrt{(k_1-k_3)^2+ 4 k_2^2 }\right), \label{eval1} \notag \\
\theta_2&=&\frac{1}{2} \left(k_1 + k_3 + \sqrt{(k_1-k_3)^2+ 4 k_2^2}\right), \label{eval2} \notag\\
\theta_3&=&\frac{2 \sqrt{2}}{\pi }+\frac{\pi }{4}-\frac{1}{\sqrt{2}},\label{eval3} \notag \\
\theta_4&=&-\frac{2 \sqrt{2}}{\pi }+\frac{\pi }{4}+\frac{1}{\sqrt{2}}.\label{eval4}
\end{eqnarray}
{ Finally, for $x=1$ the results are similar to the case of $x=0$.}

\subsection{Edge $\alpha_y=0, \alpha_z=0 $}
\label{edge2}
 
{ Here we consider the output state \eqref{output} with $ \alpha_y= \alpha_z=0 $, 
so that $\alpha_x\equiv\alpha$ becomes the parameter to estimate. 
In this case, as well as in all subsequent, the expression for the average cost function $\bar{C}(x,t,\varphi_1, \varphi_2)$ turns out to be too cumbersome to be reported. 
However, here it results not depending on $x$, and upon minimization we get 
\begin{equation}
\min \bar C=-\frac{1}{2}-\frac{8}{\pi ^2}+\frac{4}{\pi }+\frac{\pi ^2}{48} \approx 0.17,
\end{equation}
 when $ t=\frac{1}{2}$  (so again with a superposition of canonical basis states for $E$) and 
 $\varphi_1=\varphi_2=0$.
Nevertheless $\Theta$ depends on $x$ and for any $x$ Eq. \eqref{eqM} has infinitely many solutions.
For the sake of simplicity we present the optimal strategy for $x=0$, namely for}
\begin{equation}
\ket {\Psi}_{in}=\ket {1_R1_A}\left( \frac{1}{\sqrt 2} \ket{ 0_E}+ \frac{1}{\sqrt 2} \ket{ 1_E} \right ),
\end{equation}
giving the minimizing operator
\begin{equation}
\Theta=\left(
\begin{array}{cccc}
k_1 &k_2 & 0 & 0 \\
 k_2 &k_3  & 0 & 0 \\
 0 & 0 & \frac{1}{4} \left(2-\frac{8}{\pi }+\pi \right) &  \frac{i (\pi-4 )}{2 \pi } \\
 0 & 0 & -\frac{i (\pi-4 )}{2 \pi } &  \frac{1}{4} \left(-2+\frac{8}{\pi }+\pi \right) \\
\end{array}
\right),
\end{equation}
where  again $k_1, k_2, k_3$ are arbitrary real constants satisfying \eqref{k1} and \eqref{k2}.
{ Its normalized eigenvectors are}
\begin{eqnarray}
\ket \theta_1& =&\left(
\begin{array}{c}
\frac{sign(k_2)}{\sqrt{2}} \sqrt{1-\frac{k_1-k_3}{\sqrt{(k_1-k_3)^2+4 k_2^2}}}
 \\ \frac{\sqrt{2k_2^2}}{\sqrt{4k_2^2-(k_1-k_3)\left(\sqrt{(k_1-k_3)^2+4k_2^2}-k_1+k_3\right)}}\\0\\0
\end{array}
\right), \quad \notag \\
\ket \theta_2 &=&\left(
\begin{array}{c}
-\frac{sign(k_2)}{\sqrt{2}} \sqrt{1+\frac{k_1-k_3}{\sqrt{(k_1-k_3)^2+4 k_2^2}}}
 \\ \frac{\sqrt{2k_2^2}}{\sqrt{4k_2^2+(k_1-k_3)\left(\sqrt{(k_1-k_3)^2+4k_2^2}+k_1-k_3\right)}}\\0\\0
\end{array}
\right), \notag \\
\ket \theta_3 &=&\left(
\begin{array}{c}
0 \\ 0 \\-\frac{i \left(\sqrt{2}+1\right)}{\sqrt{4+2 \sqrt{2}}} \\\frac{1}{\sqrt{4+2 \sqrt{2}}}
\end{array}
\right), \quad
\ket \theta_4 =\left(
\begin{array}{c}
0 \\ 0 \\ \frac{i \left(\sqrt{2}-1\right)}{\sqrt{4-2 \sqrt{2}}} \\\frac{1}{\sqrt{4-2 \sqrt{2}}}
\end{array}
\right),
\end{eqnarray}
while the eigenvalues coincide with those in Eqs.\eqref{eval4}.

\subsection{ Edge $\alpha_x=\alpha_y, \alpha_z=0 $} 
\label{edge3}

{ In this case  we take the output state \eqref{output} with $\alpha_x=\alpha_y, \alpha_z=0 $,
so that $\alpha_x=\alpha_y\equiv\alpha$ becomes the parameter to estimate. 
Proceeding like in Sec.\ref{edge1} the minimum average quadratic cost
\begin{equation}
 \min \bar C=\frac{\pi ^2}{48 } -\frac{1}{\pi ^2 }\approx 0.10,
 \label{costedge3} 
\end{equation}
is obtained for $x=0$, $t=1$, i.e. $\ket {\Psi}_{in}=\ket {1_R1_A 0_E}$.
The minimizing operator results}
\begin{equation}
\Theta=\left(
\begin{array}{cccc}
 -\frac{1}{\pi }+\frac{\pi }{4} & 0 & 0 & 0 \\
 0 & \frac{1}{\pi }+\frac{\pi }{4} & 0 & 0 \\
 0 & 0 & k_1 & k_2 \\
 0 & 0 & k_2 & k_3 \\
\end{array}
\right),
\end{equation}
where again  $k_1, k_2, k_3$ are arbitrary real constants satisfying\eqref{k1} and \eqref{k2}.
{ Its normalized eigenvectors and eigenvalues are:}
\begin{eqnarray}
\ket \theta_1 &=&\left(
\begin{array}{c}
0\\0 \\
-\frac{sign(k_2)}{\sqrt{2}} \sqrt{\frac{k_1-k_3}{\sqrt{(k_1-k_3)^2+4 k_2^2}}+1}
 \\ \frac{\sqrt{2k_2^2}}{\sqrt{4k_2^2+(k_1-k_3)\left(\sqrt{(k_1-k_3)^2+4k_2^2}+k_1-k_3\right)}}
\end{array}
\right), \quad \notag \\
\ket \theta_2 &=&\left(
\begin{array}{c}
0 \\ 0 \\ \frac{sign(k_2)}{\sqrt{2}} \sqrt{1-\frac{k_1-k_3}{\sqrt{(k_1-k_3)^2+4 k_2^2}}}
 \\ \frac{\sqrt{2k_2^2}}{\sqrt{4k_2^2-(k_1-k_3)\left(\sqrt{(k_1-k_3)^2+4k_2^2}-k_1+k_3\right)}}
\end{array}
\right), \notag \\
\ket \theta_3 &=&\left(
\begin{array}{c}
0 \\ 1 \\ 0\\ 0
\end{array}
\right),  \quad
\ket \theta_4 =\left(
\begin{array}{c}
1 \\ 0 \\ 0\\ 0
\end{array}
\right),
\end{eqnarray}
and
\begin{eqnarray}
\theta_1&=&\frac{1}{2} \left(-\sqrt{(k_1- k_3)^2+4 k_2^2}+k_1+k_3\right), \notag  \\
\theta_2&=&\frac{1}{2} \left(\sqrt{(k_1- k_3)^2+4 k_2^2}+k_1+k_3\right),  \notag  \\
\theta_3&=&\frac{1}{\pi }+\frac{\pi }{4}, \notag  \\
\theta_4&=&-\frac{1}{\pi }+\frac{\pi }{4}.
\end{eqnarray}
The minimum average quadratic cost \eqref{costedge3} can be also achieved with a POVM characterized by the above eigenvectors and eigenvalues together with the input 
\begin{equation}
\ket {\Psi}_{in}=\ket {0_R0_A 1_E}. 
\end{equation}
\subsection{Edge $\alpha_x=\alpha_y=\alpha_z$ }
\label{edge4}

{ When $\alpha_x=\alpha_y=\alpha_z$
(so that $\alpha_x=\alpha_y=\alpha_z\equiv\alpha$ becomes the parameter to estimate)
 the results (minimum average cost function, optimal POVM and optimal input state) are the same as in Sec.\ref{edge3}.
}


\subsection{Edge $ \alpha_x=\frac{\pi}{2}, \alpha_z=0$}
\label{edge5}

{ Here we consider the output state \eqref{output} with $ \alpha_x=\frac{\pi}{2}, \alpha_z=0$
so to have $\alpha_y\equiv\alpha$ as parameter to estimate.
Upon minimization of the average cost function $\bar{C}(x,t,\varphi_1, \varphi_2)$ we get 
\begin{equation}
 \min \bar C=-\frac{1}{2}-\frac{8}{\pi ^2}+\frac{4}{\pi }+\frac{\pi ^2}{48}\approx 0.17,
\end{equation}
when $x=\frac{1}{2}$, $t=1$ (or equivalently $t=0$) and $\varphi_1=\varphi_2=0$, i.e.
\begin{equation}
\ket {\Psi}_{in}=\frac{1}{\sqrt 2}\left( \ket {0_R 0_A}+\ket {1_R 1_A} \right ) \ket 0_E.
\end{equation}
In this case a maximally entangled input gives the smallest average quadratic cost.}

The minimizing operator turns out to be
\begin{equation}
\Theta=\left(
\begin{array}{cccc}
 \frac{1}{4} \left(-2+\frac{8}{\pi }+\pi \right) & 0 & 0 & \frac{1}{2}-\frac{2}{\pi } \\
 0 & \frac{1}{4} \left(2-\frac{8}{\pi }+\pi \right) & \frac{1}{2}-\frac{2}{\pi } & 0 \\
 0 & \frac{1}{2}-\frac{2}{\pi } & \frac{1}{4} \left(-2+\frac{8}{\pi }+\pi \right) & 0 \\
 \frac{1}{2}-\frac{2}{\pi } & 0 & 0 & \frac{1}{4} \left(2-\frac{8}{\pi }+\pi \right) \\
\end{array}
\right).
\end{equation}
Its normalized eigenvectors and eigenvalues are
\begin{eqnarray}
\ket \theta_1 &=&\left(
\begin{array}{c}
\\-\frac{\sqrt{2}+1}{\sqrt{2 \left(\sqrt{2}+2\right)}} \\ 0 \\ 0 \\ \frac{1}{\sqrt{2 \left(\sqrt{2}+2\right)}}
\end{array}
\right),\hspace{7mm}
\ket \theta_2 =\left(
\begin{array}{c}
0 \\\frac{1-\sqrt{2}}{\sqrt{2\left(2- \sqrt{2}\right)}} \\ \frac{1}{\sqrt{2\left(2- \sqrt{2}\right)}} \\ 0
\end{array}
\right),\hspace{5mm} \notag \\
\ket \theta_3 &=&\left(
\begin{array}{c}
\frac{\sqrt{2-\sqrt{2}}}{2} \\ 0 \\ 0 \\ \frac{\sqrt{2-\sqrt{2}}\left(\sqrt{2}+1\right)}{2}
\end{array}
\right),\quad 
\ket \theta_4 =\left(
\begin{array}{c}
0 \\ \frac{\sqrt{2+\sqrt{2}}}{2} \\ \frac{\sqrt{2+\sqrt{2}}\left(1-\sqrt{2}\right)}{2}\\ 0
\end{array}
\right),
\end{eqnarray}
and 
\begin{eqnarray}
\theta_1&=&\theta_2=\frac{2 \sqrt{2}}{\pi }+\frac{\pi }{4}-\frac{1}{\sqrt{2}}, \notag \\
\theta_3&=&\theta_4=-\frac{2 \sqrt{2}}{\pi }+\frac{\pi }{4}+\frac{1}{\sqrt{2}}.
\end{eqnarray}

\subsection{Edge $\alpha_x=\frac{\pi}{2},\alpha_y=\alpha_z $} 
\label{edge6}

{ At the end we set $\alpha_x=\frac{\pi}{2},\alpha_y=\alpha_z $ in the output state \eqref{output}
so to have $,\alpha_y=\alpha_z \equiv\alpha$ as parameter to estimate.
The quantity $\bar{C}(x,t,\varphi_1, \varphi_2)$ attains the minimum
\begin{equation}
\min \bar C =\frac{128+256 \pi -344 \pi ^2+128 \pi ^3-24 \pi ^4+\pi ^6}{48 \pi ^2 \left(\pi ^2-8\right)} \approx 0.15,
\end{equation}
when $x=t=\frac{1}{2}$ and $\varphi_1=\varphi_2=0$, i.e. for 
}
\begin{equation}
\ket {\Psi}_{in}=\frac{1}{\sqrt 2}\left( \ket {0_R0_A}+\ket {1_R1_A} \right ) \left( \frac{1}{\sqrt 2} \ket 0_E+ \frac{1}{\sqrt 2} \ket 1_E \right ).
\end{equation}
{ In this case maximally entangled input and superposition of canonical basis states of environment give the smallest average quadratic cost.}

The minimizing operator results
\begin{equation}
\Theta=\left(
\begin{array}{cccc}
 k & \mu-k & \gamma+k &\sigma -k \\
 \mu^*-k & k &\sigma -k & \gamma^*+k \\
 \gamma^*+k & \sigma -k & k & \mu^*-k \\
\sigma -k & \gamma+k & \mu-k & k \\
\end{array}
\right),
\end{equation}
where
\begin{eqnarray}
\mu&=&\frac{-64+(32+8 i) \pi -(26+16 i) \pi ^2+4 i \pi ^3+3 \pi ^4}{12 \pi  \left(\pi ^2-8\right)},\\
\gamma&=& \frac{32-(64-8 i) \pi +(40-16 i) \pi ^2+4 i \pi ^3-3 \pi ^4}{12 \pi  \left(\pi ^2-8\right)},\\
\sigma&=&\frac{32-18 \pi +\pi ^3}{4 \left(\pi ^2-8\right)},
\end{eqnarray}
and  $k$ is an arbitrary real constant satisfying
\begin{equation}
\frac{-32+64 \pi -40 \pi ^2+3 \pi ^4}{16 \pi ^3-128 \pi }\leq k\leq \frac{-32+64 \pi -56 \pi ^2+5 \pi ^4}{16 \pi ^3-128 \pi }.
\end{equation}
{ Again, this constraint is imposed by the fact that the parameter we are estimating must lie in the interval 
$\left[0, \frac{\pi}{2}\right]$ and hence must do the eigenvalues of minimizing operator.}
{ Due to the complexity of solutions, here we present the eigenvectors and eigenvalues of $\Theta$
in numerical (approximated) form:}
\begin{eqnarray}
\ket \theta_1&=&\left( \begin{array}{c} -0.5\\-0.5\\0.5\\0.5 \end{array} \right), \quad
\ket \theta_2=\left( \begin{array}{c} -0.5\\0.5\\-0.5\\0.5 \end{array} \right), \notag \\
\ket \theta_3&=&\frac{sign(k-0.99)}{\sqrt{k^2-1.98 k+0.98}}
\left( \begin{array}{c} 
0.5(k-0.99)\\
(-0.15-0.47 i)+(0.15 +0.47 i) k\\
(-0.15-0.47 i)+(0.15 +0.47 i) k\\
0.5(k-0.99)
\end{array} \right), \notag
\\
\ket \theta_4&=&\frac{sign(k-0.86)}{\sqrt{k^2-1.72 k+0.74}}
\left( \begin{array}{c} 
0.5(k-0.86)\\
(0.13+0.41 i)-(0.15+0.48 i) k\\ 
(0.13+0.41 i)-(0.15+0.48 i) k\\ 
0.5(k-0.86)
\end{array} \right), 
\end{eqnarray}
and
\begin{eqnarray}
\theta_1=1.10,\quad
\theta_2=-2.83+4 k,\quad
\theta_3=1.12,\quad
\theta_4=0.60.
\end{eqnarray}


\section{Conclusion}

{ In conclusion we have considered the problem of estimating a parametrized family of system-environment interaction unitaries by analysing the channels arising from it. To this end we have employed a model of quantum channel with environment assistance. 
In practice we have focussed on two qubit unitaries when initial states of both qubit can be locally controlled, while the final state of only one qubit can be measured. 
In such a framework the parameters space results a tetrahedron in $\mathbb{R}^3$.  
On its edges we have found optimal probe states and estimators (POVM) 
 by minimizing the average quadratic cost function. }
  
Those optimal strategies we have found can be divided into three types:
\begin{itemize}
\item 
Control of both probe and environment states is needed (probe must be maximally entangled with  reference system): edges $\alpha_x=\frac{\pi}{2}, \alpha_z=0 $ and $\alpha_x=\frac{\pi}{2},\alpha_y=\alpha_z $; 
\item  
Control of both probe state and environment state is needed (probe must be factorable with reference system): 
edges $\alpha_x=\alpha_y=\alpha_z $  and  $\alpha_x=\alpha_y, \alpha_z=0 $;
\item 
Control of only environment state is needed: 
edges $ \alpha_x=\frac{\pi}{2}, \alpha_y=\frac{\pi}{2} $ and $ \alpha_y=0, \alpha_z=0 $.
\end{itemize}
In practice this shows that the possibility of environment assistance is always beneficial, while the usage of entanglement at channels input is not. Even more, controlling the probe state is not always necessary. 

Although the optimal strategies were derived with input state having the form of Eq.\eqref{input}, 
it is possible to show that the same values of minimum average quadratic cost can be attained 
by means of the input
\begin{equation}
\left(\sqrt{x}\ket {0_R} \ket {1_A}+e^{i\varphi_1}\sqrt{1-x}\ket{ 1_R} \ket {0_A}\right) \left(\sqrt{t}\ket {0_E} 
+e^{i\varphi_2}\sqrt{1-t}\ket {1_E} \right),
\end{equation}
with suitable values of $x, t,\varphi_1,\varphi_2$ and POVMs.

Actually the lowest value for the minimum average cost function ($\approx0.10$) is achieved on the edges 
($\alpha_x=\alpha_y$, $\alpha_z=0$) and ($\alpha_x=\alpha_y=\alpha_z$) meaning that there the parameter can be better estimated than in the other edges of the tetrahedron.

The present work paves the way for several future developments.
Clearly it is desirable to solve the problem also inside the tetrahedron \eqref{spaceS}.
Here the difficult comes from the multi-parameter estimation for which there 
are not known optimization algorithms. Perhaps the usage of local strategies 
could be useful to this end. These amount to look for a POVM maximizing the
Fisher information, thus minimizing the variance of the estimator, at fixed values of 
parameters \cite{Helstrom}. 
It is also foreseeable an extension to higher dimensional unitaries ${\rm SU}(d\times d)$ and even to infinite dimensional systems by restricting to Gaussian systems and resorting to symplectic representations \cite{Weed}.
Additionally, one can consider unitaries where environment is larger than the main system, 
given the fact that a channel of dimension $d$ can always be dilated to comprise an environment of dimension at most $d^2$ \cite{mark}.

\section*{Acknowledgments}

The work of M.R. is supported by China Scholarship Council.

\end{document}